\documentclass[aps,twocolumn,showpacs,amsmath,amssymb,superscriptaddress,prl]{revtex4-1}
\usepackage{graphicx}
\usepackage{dcolumn}
\usepackage{bm}
\usepackage{gensymb}

\newcommand{\rowgroup}[1]{\hspace{-1em}#1}
\providecommand{\e}[1]{\ensuremath{\times 10^{#1}}}

\begin{document}

\title{Precision Measurement of the Radiative $\beta$ decay of the Free Neutron}

\author{M.J.~Bales}\email[Corresponding author: ]{matthew.bales@tum.de}\affiliation{University of Michigan, Ann Arbor, Michigan 48104, USA}\affiliation{Physikdepartment, Technische Universit{\"a}t M{\"u}nchen, D-85748, Germany}
\author{R.~Alarcon}\affiliation{Arizona State University, Tempe, Arizona 85287, USA}
\author{C.D.~Bass}\altaffiliation[Present Address: ]{Le Moyne College, Syracuse, New York 13214, USA}\affiliation{National Institute of Standards and  Technology, Gaithersburg, Maryland 20899, USA}
\author{E.J.~Beise}\affiliation{University of Maryland, College Park, Maryland 20742, USA}
\author{H.~Breuer}\affiliation{University of Maryland, College Park, Maryland 20742, USA}
\author{J.~Byrne}\affiliation{University of Sussex, Brighton, BN1 9QH, United Kingdom}
\author{T.E.~Chupp}\affiliation{University of Michigan, Ann Arbor, Michigan 48104, USA}
\author{K.J.~Coakley}\affiliation{National Institute of Standards and  Technology, Boulder, Colorado 80305, USA}
\author{R.L.~Cooper}\altaffiliation[Present Address: ]{New Mexico State University, Las Cruces, New Mexico 88003-8001, USA}\affiliation{Indiana University, Bloomington, Indiana 47408, USA}
\author{M.S.~Dewey}\affiliation{National Institute of Standards and  Technology, Gaithersburg, Maryland 20899, USA}
\author{S.~Gardner}\affiliation{University of Kentucky, Lexington, Kentucky 40506 USA}
\author{T.R.~Gentile}\affiliation{National Institute of Standards and  Technology, Gaithersburg, Maryland 20899, USA}
\author{D.~He}\altaffiliation[Present Address: ]{Center for High Energy Physics, Peking University, Beijing 100871,
China}\affiliation{University of Kentucky, Lexington, Kentucky 40506 USA}
\author{H.P.~Mumm}\affiliation{National Institute of Standards and  Technology, Gaithersburg, Maryland 20899, USA}
\author{J.S.~Nico}\affiliation{National Institute of Standards and  Technology, Gaithersburg, Maryland 20899, USA}
\author{B.~O'Neill}\affiliation{Arizona State University, Tempe, Arizona 85287, USA}
\author{A.K.~Thompson}\affiliation{National Institute of Standards and  Technology, Gaithersburg, Maryland 20899, USA}
\author{F.E.~Wietfeldt}\affiliation{Tulane University, New Orleans, Louisiana 70118, USA}

\collaboration{RDK II Collaboration}\noaffiliation
\date{\today}

\begin{abstract}
The standard model predicts that, in addition to a proton, an electron, and an antineutrino, a continuous spectrum of photons is emitted in the $\beta$ decay of the free neutron.  We report on the RDK II experiment which measured the photon spectrum using two different detector arrays.  An annular array of bismuth germanium oxide scintillators detected photons from 14 to 782~keV.  The spectral shape was consistent with theory, and we determined a branching ratio of 0.00335 $\pm$ 0.00005 [stat] $\pm$ 0.00015 [syst]. A second detector array of large area avalanche photodiodes directly detected photons from 0.4 to 14~keV. For this array, the spectral shape was consistent with theory, and the branching ratio was determined to be 0.00582 $\pm$ 0.00023 [stat] $\pm$ 0.00062 [syst].  We report the first precision test of the shape of the photon energy spectrum from neutron radiative decay and a substantially improved determination of the branching ratio over a broad range of photon energies.
\end{abstract}



\maketitle{}

In the six decades since the first measurement of the neutron lifetime, the study of neutron beta decay has provided increasingly precise tests of the standard model and important input to cosmology and other areas of physics~\cite{Abele08, Nico09, Dubbers11}. Precision measurements of neutron observables, such as the lifetime~\cite{Wietfeldt11,pdg2015} and the spin-electron asymmetry coefficient~\cite{Mund13, Mendenhall13, pdg2015}, allow for comparisons with theory with a precision below 1\%.  The standard model predicts that the decay of the free neutron can produce one or more detectable radiative photons in addition to a proton, an electron, and an antineutrino.  Calculated radiative corrections of approximately 4\% are employed in relating the measured lifetime to weak interaction parameters~\cite{Marciano2006}.   Given the precision of neutron beta-decay measurements,  it is important to perform direct precision measurements of its radiative decay mode.    

Here we present the results of the RDK II experiment, which includes the first precision test of the shape of the photon energy spectrum and a substantially improved determination of the branching ratio. This demonstrates the feasibility of precise measurements of the neutron's radiative decay mode that can probe additional physics.  For example, a measurement of the photons' circular polarization could reveal information about the Dirac structure of the weak current~\cite{gaponov1, bernard1, robthesis} and a possible source of time-reversal violation would be apparent in a triple-product correlation between the antineutrino, electron, and photon~\cite{gardner1, gardner2}. Increased precision would allow a test of a heavy baryon chiral perturbation theory calculation~\cite{bernard1}.

In contrast with the long history of neutron beta-decay measurements, experimental studies of neutron radiative beta decay are relatively recent. An experiment in 2002 placed a limit on the branching ratio for this process~\cite{neutronraddecayexp1}, and in 2006 the RDK I collaboration reported the first definitive observation of radiative decay~\cite{rdkpaper1,rdkpaper2}.  The RDK II experiment~\cite{rdk2det, benthesis, balesThesis,rdkpaper2} improved upon its predecessor by reducing statistical uncertainties through the use of additional photon detectors, improving the understanding of systematic uncertainties through detailed energy response studies of the detectors, and significantly extending the detectable photon energy range to between 0.4 keV and the 782 keV photon energy end point.

Radiative photons from neutron decay originate from either  electron,  proton, or vertex bremsstrahlung.  Electron bremsstrahlung dominates while the recoil order terms, including vertex bremsstrahlung, contribute less than 1\% to the branching ratio~\cite{bernard1}. We performed our own numerical calculations using leading order QED~\cite{robthesis} without accounting for finite-nucleon-size effects.  Our calculations agree with branching ratios from other published calculations~\cite{gaponov1, ivanov1, bernard1} to within 1\%.  We used (880.3 $\pm$ 1.1) s for the neutron lifetime~\cite{pdg2015} and included a Coulomb correction of 3\% to the radiative partial decay rate, which was not present in prior calculations~\cite{Wilkinson1982,HeThesis}.  Other next-to-leading order effects were not included~\cite{HeThesis}.

The experiment operated at the NG-6 fundamental physics end station at the Center for Neutron Research (NCNR) at the National Institute of Standards and Technology (NIST)~\cite{ng6beamline}.  The reactor-produced cold neutron beam was guided to the experiment as in RDK I~\cite{rdkpaper2}, but with increased collimation to decrease backgrounds and systematic uncertainties in decay locations. Using a calibrated \textsuperscript{6}Li-foil neutron flux monitor~\cite{nicolifetime, yue2013} mounted downstream of the detection region, the typical neutron rate was determined to be 1.1\e{8} /s.

\begin{figure}
 \includegraphics[width=0.49\textwidth]{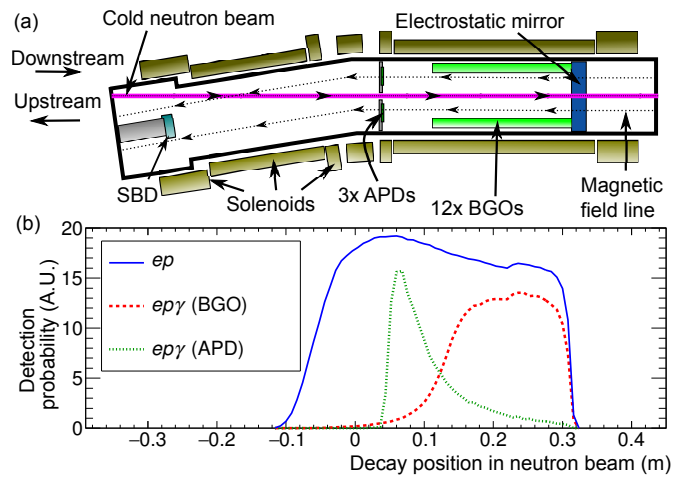}
 \caption
 	{\label{fig:RDKIIDiagram}
 	(a) A cross sectional diagram of the RDK II detection apparatus from above. The neutron beam (pink) traveled from left to right through the active region defined by the fields (dashed lines) created by the solenoids (gold) and the electrostatic mirror (blue). Protons and electrons follow the field lines to the SBD (light blue).  Radiative photons are detected by twelve BGO crystals (green) and three large area APDs.\\
(b) Detection probability in independent arbitrary units (A.U.) for electron-proton ($ep$) and electron-proton-photon ($ep\gamma$) detection coincidence for either the BGO or direct APD detectors. This plot is approximately aligned with the diagram above.
 	}
 \end{figure}

The neutron beam passed through a strong magnetic field produced by a set of superconducting solenoids that were used to guide charged decay products to a detector. This detection method has been used in several experiments measuring neutron decay parameters~\cite{Byrne1990,Byrne2002,Dewey2003,rdkpaper1}. The detection region  [see Fig. \ref{fig:RDKIIDiagram} (a)] was defined by a 9.5\degree{} bend in the magnetic field and a ring of aluminum maintained at +1400 V that served as an electrostatic mirror. The mirror created an +800 V barrier at the center of the beam to protons.  The magnetic field varied from 3.3 to 4.6 T over the 34 cm distance between the bend and mirror.

Neutrons which decayed between the mirror and the bend produced electrons and protons capable of being detected by the experiment.  The electrons and protons followed adiabatic helical orbits about the field lines with maximum cyclotron radii of approximately 1~mm.  Decay electrons have typical kinetic energies of hundreds of keV.  Electrons emitted in the upstream direction followed the magnetic field to a 1 or 1.5~mm thick, 600~mm\textsuperscript{2} silicon surface barrier detector (SBD) in a time on the order of nanoseconds, whereas those emitted in the the downstream direction [see Fig. \ref{fig:RDKIIDiagram}(a)] typically escaped the active detection region undetected. Protons were detected if emitted in either direction because the electrostatic mirror was sufficient to reflect all of them.  The protons traveled to the SBD in a time on the order of microseconds.  The SBD was held at a $-$25~kV potential to accelerate protons through the gold layer on its front face.  The SBD was calibrated by determining the electron endpoint energy of neutron decay from a functional fit, and its linearity was verified with radioactive source measurements.

Two separate photon detector arrays surrounded the neutron beam in the active region.  The BGO detector array consisted of twelve 1.2~cm x 1.2~cm x 20~cm bismuth germanium oxide (BGO) scintillator crystals optically coupled to avalanche photodiodes (APDs)~\cite{rdk2det}.  The detection range of the BGO detectors was between approximately 10 to 1000~keV.  The cryogenic environment (80~K) inside the detector served to both increase the BGO scintillators' light output and the APDs' gain while decreasing the APDs' noise.  The 12 detectors were, for the most part, shielded from bremsstrahlung associated with particles striking the SBD. A small correction and uncertainty for this process was determined by simulation (see Table \ref{table:Systematics}).

The typical BGO detector energy resolution was 10\% (full width at half maximum) at 662~keV and  30\% at 60~keV~\cite{rdk2det}.  When the neutron beam shutter was open, a 511~keV electron-positron annihilation peak was observed in the BGO photon background spectrum and was used for calibration, i.e., photon pulse height channel 511 corresponds to 511~keV.  The nonproportionality of light output versus photon energy deposited in BGO crystals is significant at lower energies and caused the BGO lower limit at photon pulse height channel 10 to be centered at 14.1~keV of photon energy deposited. This nonproportionality was measured in a separate study~\cite{Gentile2015} and parametrized by a consensus model in combination with existing literature~\cite{khodyuk, moszynski, verdier, sysoeva, averkiev}.  This model was then incorporated into the Monte Carlo (MC) simulation.

BGO nonproportionality was not accounted for in the RDK I experiment.  In addition, the theoretical value quoted in Refs.~\cite{rdkpaper1,rdkpaper2} did not include a Coulomb correction to the outgoing decay electron~\cite{Wilkinson1982, HeThesis}; this should be the largest effect at order $\alpha^2$, where $\alpha$ is the fine structure constant.  Incorporating both of these changes would not alter the value of the previously measured branching ratio, 0.00309 $\pm$ 0.00032, but would  adjust the RDK I energy range to 19--313~keV instead of 15--340~keV and result in a corrected theoretical branching ratio of 0.00259.

The APD array consisted of three 2.8~cm x 2.8~cm APDs that directly detected 0.4 to 14~keV photons without a scintillator~\cite{rdk2det}.  The detection range of the APD detectors was between approximately 0.3 to 20~keV. The APDs were oriented with their bias field parallel to the magnetic field due to previously reported issues with x-ray detection if the APDs were oriented with their bias field perpendicular to the magnetic field at low temperature~\cite{rdk2bapdmagfield}. During data taking, the APDs were exposed to a \textsuperscript{55}Fe radioactive source mounted near the detectors, which produced 5.9~keV photons for calibration.  Off-line studies at synchrotron sources were performed to explore the APDs' complex energy response, which is due to reduced charge collection efficiency for photons that are absorbed in the front 1 $\mu$m of the APD~\cite{rdk2bapdresponse}.  Models of the charge collection efficiency of the APDs' doped layers of Si were created and then incorporated into the MC simulation of the detectors.

Data recording~\cite{rdk2det,benthesis} was triggered by two single channel analyzers and a time-to-amplitude converter (TAC). An SBD signal equivalent to $>$ 50~keV (an electron) followed by a signal equivalent to $>$ 7~keV (an accelerated proton), with both falling within the 25~$\mu$s time range of the TAC, initiated data recording to disk from the SBD and both photon detector arrays. The waveforms of all signals were recorded from 25~$\mu$s before to 57~$\mu$s after the electron signal with 2048 channel resolution.

\begin{figure}
 \includegraphics[width=0.49\textwidth]{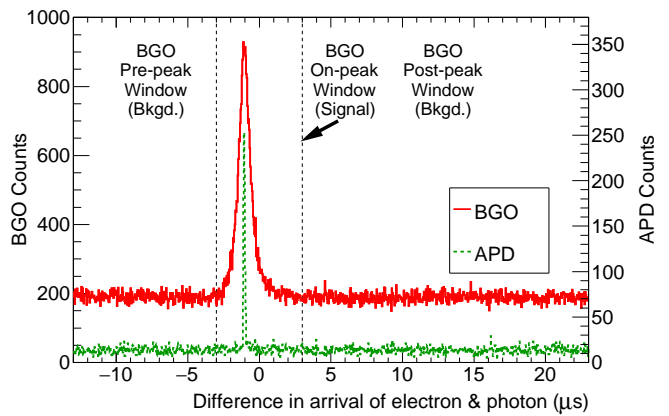}
 \caption
 	{\label{fig:egTime}
		The detected timing spectrum for the difference between electron and photon detection in coincidence with a delayed proton.  The central peak arises from radiative photons which are detected nearly simultaneously with the electrons while the flat regions represent sources of constant, uncorrelated photon background.  The response of the APD detectors was significantly faster than the BGO detectors and resulted in a sharper timing peak.  Only the background and signal windows for the BGO detectors are shown.
 	}
 \end{figure}

The RDK II experiment operated on the neutron beam line from December 2008 until November 2009. The final data set consisted of 22 million electron-proton ($ep$) detections, for which about 20\,000 and 800 radiative photons were detected in  coincidence ($ep\gamma$) with the BGO and APD detectors, respectively. Some data runs were eliminated from the analysis typically for one of three reasons: their small size, a loss of SBD detector gain, or an anomalous structure in the proton time-of-flight spectrum that was not consistent with simulation.  The waveform data were analyzed and the waveforms from the SBD and the photon detectors were fit to functional forms to extract their pulse heights.  For each individual photon detector,  background was determined by using the prepeak and postpeak photon backgrounds found in the electron-photon timing spectrum (see Fig.~\ref{fig:egTime}).

To account for the experiment's complex detection efficiency profile [see Fig. \ref{fig:RDKIIDiagram} (b)], a MC simulation was created.  In the simulation, initial momenta and positions for the neutron decay products were created randomly using a leading order QED event generator~\cite{robthesis} and a simulation of the neutron beam profile.  The protons, electrons, and photons were then transported by a Runge-Kutta algorithm in a model of the detection region's geometry with \textsc{geant4.9.6.p02}~\cite{geant4, balesThesis}.  Magnetic and electric fields were interpolated from simulated field maps of the apparatus. \textsc{geant4.9.6.p02} was also used to determine the energy deposited in the detectors, including any secondary radiation or backscattering produced.  Models of detector energy response and energy resolution were also incorporated. Results from the simulations were consistent with experimental timing and energy spectra of the protons and electrons.

\begin{figure*}
 \includegraphics[width=0.99\textwidth]{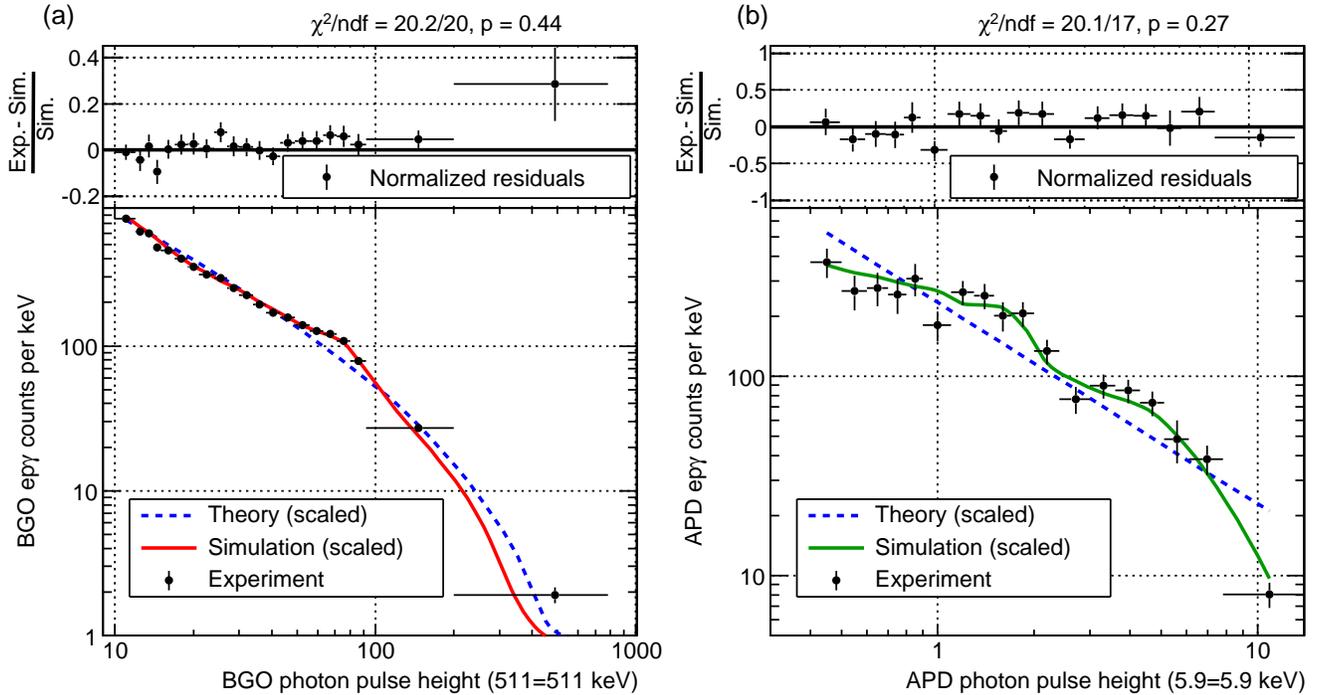}
 \caption
 	{\label{fig:spectra}
		Energy spectrum deposited by photons from radiative neutron decay.  Plotted are the average background-corrected radiative photon counts for both BGO, (a), and APD, (b), detectors versus photon pulse height. The pulse height was scaled such that it is approximately equal to the photon energy deposited. The blue dashed line shows the theoretical spectrum scaled to the experimental data and plotted versus photon energy. The solid lines are the output of the simulation scaled to the experimental data using the theoretical spectrum as input. The simulation incorporates the coincident detection of the decay particles and the response functions for the BGO array (red) and the APD array (green). The bump seen in (a) at $\approx$80 keV is caused by the escape of bismuth K x-rays from nearby crystals.  The experimental data (black circles), include only the statistical uncertainty in the vertical error bars while the horizontal error bars represent the bin size.  The normalized residuals (black circles) between the experiment and simulation are also shown at the top along with the results of a $\chi^2$ evaluation. Here ndf is the number of degrees of freedom and $p$ is the $p$ value.
 	}
 \end{figure*}

A ratio was formed for both the experimental data and the simulated data by dividing the detected rate of electron, proton, and photon coincidences $r_{ep\gamma}$ by the detected rate of electron and proton coincidences $r_{ep}$.  This ratio of rates $R = r_{ep\gamma} / r_{ep}$ serves two purposes: it is independent of the neutron rate and many systematic uncertainties associated with the electron and proton detection cancel.  The ratio for the integrated experimental data $R_{\textrm{expt}}$  and for the integrated simulated data $R_{\textrm{sim}}$ can be then compared. Because $R_{\textrm{sim}}$ is dependent on the theoretical branching ratio $B_{\textrm{theory}}$, an experimental branching ratio can be extracted $B_{\textrm{expt}}=B_{\textrm{theory}} R_{\textrm{expt}} / R_{\textrm{sim}}$.

 \begin{table}
 \scalebox{.95}{
 \tabcolsep=0.07cm
 \footnotesize
\begin{ruledtabular}
\begin{tabular}{>{\quad}lcccc}
 \footnotesize  			& \footnotesize BGO& \footnotesize BGO& \footnotesize APD& \footnotesize APD\\
 \footnotesize  			& \footnotesize Corr.(\%)& \footnotesize Unc.(\%)& \footnotesize Corr.(\%)& \footnotesize Unc.(\%)\\
\hline
\rowgroup{Photon detectors}		& &	& &	\\
Energy response					& $\cdot\cdot\cdot$		& 2.6	& $\cdot\cdot\cdot$		& 10	\\
Photon energy calibration		& $\cdot\cdot\cdot$		& 0.6	& $\cdot\cdot\cdot$		& 1.2		\\
Multiple photons				& 0.4	& 0.2	& 0.1	& 0.1		\\
\rowgroup{SBD detector}			& & & &	\\
Electron energy calibration		& $\cdot\cdot\cdot$		& 0.2	& $\cdot\cdot\cdot$		& 0.5		\\
Proton energy calibration		& $\cdot\cdot\cdot$		& 0.5	& $\cdot\cdot\cdot$		& 0.4		\\
Pulse shape discrimination		& $\cdot\cdot\cdot$		& 2.2	& $\cdot\cdot\cdot$		& 0.4		\\
\rowgroup{Timing cuts}		 	& &	& &	\\
Electron-proton timing  		& $\cdot\cdot\cdot$		& 0.5	& $\cdot\cdot\cdot$		& 0.6		\\
Electron-photon timing			& $\cdot\cdot\cdot$		& $\cdot\cdot\cdot$		& $\cdot\cdot\cdot$		& $\cdot\cdot\cdot$		\\
\rowgroup{$ep\gamma$ backgrounds}& & & &	\\
Electron bremsstrahlung			& -0.8	& 0.1	& $\cdot\cdot\cdot$		& $\cdot\cdot\cdot$		\\
Non-decay background 		& -1.0	& 0.8	& -0.4	& 0.4		\\
\rowgroup{Simulation}	& &	& &	\\
Model registration          	& $\cdot\cdot\cdot$		& 2.7	& $\cdot\cdot\cdot$		& 3.6    \\
Statistics						& $\cdot\cdot\cdot$		& 0.1	& $\cdot\cdot\cdot$		& 0.4		\\
\hline
\rowgroup{Total Systematic} 	& -1.4	& 4.7	& -0.3	& 11		\\
\end{tabular}
\end{ruledtabular}
}
\caption{\label{table:Systematics} 
 Summary of the systematic corrections and relative standard uncertainties in the measured branching ratio for bismuth germanium oxide (BGO) and avalanche photodiode (APD) detectors.  ``$\cdot\cdot\cdot$" indicates less than 0.05\% in magnitude. 
}
\end{table}

The corrections and relative standard uncertainties of systematic effects associated with both the BGO and APD measurements are given in Table~\ref{table:Systematics}.  The dominant systematic uncertainties in this experiment were in the simulation's model registration, pulse shape discrimination, and photon detector energy response.  Bremsstrahlung induced by electrons interacting with the SBD were found in the simulations to produce photons that could mimic a radiative decay signal in the BGO detectors, and a -0.8\% correction was made.  A -1.0\% correction from nondecay coincident events was made for the BGO detectors.

	Model registration refers to the positional accuracy of the simulation's model with respect to the apparatus including the uncertainty in position of each detector.  This uncertainty resulted primarily from decays originating from the intersection of the neutron beam with the region where the magnetic field bends towards the SBD [see Fig. \ref{fig:RDKIIDiagram}(a)], for which a portion of the electrons and protons would strike the edges of the SBD or would miss the detector entirely.  Uncertainties in the position of either the neutron beam or the SBD therefore affect the simulated $r_{ep}$.  However, as seen in Fig. \ref{fig:RDKIIDiagram}(b), this region does not contribute to $r_{ep\gamma}$ so the uncertainty fails to cancel in the ratio $R_{\textrm{sim}}$.  The uncertainty was determined by varying the position of the beam and SBD in the simulation within the uncertainties of their positions, which were between 1 to 2 mm after thermal contraction.

	The pulse shape discrimination uncertainty arose from the difficulty in identifying proton pulses which occurred after some electron pulses that exhibited a slow signal decay or increased noise between the pulses.  This caused the proton pulse to be superimposed on the tail of these electron pulses, which made particle identification difficult.  The uncertainty was determined by varying an analysis cut based on the identification of these problematic electron pulses.

	The BGO energy response uncertainty was calculated from the observed variability between different experimental measurements of nonproportionality in BGO scintillators from the literature and from our crystals~\cite{Gentile2015}.  The APD energy response was evaluated from models of the electron collection efficiency versus x-ray absorption depth that were based on measurements performed with one APD using both monochromatic x-ray beams and broadband synchrotron radiation~\cite{rdk2bapdresponse}. The uncertainty arose primarily from the difference observed in the branching ratio for these two approaches.  Additional uncertainty was incorporated to account for differences observed between the three APDs.

	The radiative spectra from each BGO detector were averaged. This averaged spectrum [see Fig.~\ref{fig:spectra}(a)] agreed well with the scaled average spectrum predicted by simulation and resulted in a chi squared per degree of freedom of 20.2/20 with a $p$ value of 0.44.  The branching ratio $B_{\textrm{expt}}^{\textrm{BGO}}$ was measured to be 0.00335 $\pm$ 0.00005 [stat] $\pm$ 0.00015 [syst] in the range of 14.1 to 782 keV.  For this range, our $B_{\textrm{theory}}^{\textrm{BGO}}$ was calculated to be 0.00308. The values  $B_{\textrm{expt}}^{\textrm{BGO}}$ and $B_{\textrm{theory}}^{\textrm{BGO}}$ agree within  $1.7$ times the combined standard uncertainty.  

	The radiative spectra from each APD detector were averaged. This averaged spectrum [see Fig.~\ref{fig:spectra}(b)] agreed well with the scaled average spectrum predicted by simulation and resulted in a chi-squared per degree of freedom of 20.1/17 with a $p$ value of 0.27. The branching ratio $B_{\textrm{expt}}^{\textrm{APD}}$was measured to be 0.00582 $\pm$ 0.00023 [stat] $\pm$ 0.00062 [syst] in the range of 0.4 to 14~keV. For this range, our $B_{\textrm{theory}}^{\textrm{APD}}$ was calculated to be 0.00515. The values $B_{\textrm{expt}}^{\textrm{APD}}$ and $B_{\textrm{theory}}^{\textrm{APD}}$ agree within $1.0$ times the combined standard uncertainty.  
	
	In summary, we have reported the first precise measurement of the radiative decay of the free neutron spanning 3 orders of magnitude in photon energy using two different detectors. As the precision is limited by systematic effects, the significantly better understanding of these effects obtained in this experiment provides a path towards an improved experiment with an uncertainty below 1\%.  A future experiment could be considered that eliminates the magnetic field, which would allow for particle tracking and improved detection-volume definition.    In addition,  photon detectors with better proportionality could be implemented, and improvements in low-energy proton detection would allow better identification of proton and electron events. Utilizing a higher intensity cold neutron source should significantly improve the ability to study systematics while maintaining high statistical precision.

\begin{acknowledgments}
We thank Changbo Fu for his initial simulation work and David Winogradoff for his calibration studies.  We additionally thank R. Farrell for numerous discussions about APD operation and their properties. This research was supported in part through computational resources and services provided by Advanced Research Computing at the University of Michigan, Ann Arbor. We acknowledge the support of the National Institute of Standards and Technology, U.S. Department of Commerce, in providing the neutron facilities used in this work. This research was made possible in part by support from the National Science Foundation (Grants No. PHY-0969654, No. PHY-1205266, No. PHY-1205393, No. PHY-1306547, and No. PHY-1505196) and the U.S. Department of Energy (Grant No. DE-FG02-96ER40989 and an interagency agreement).

\end{acknowledgments}

%

\end{document}